\documentclass{article}
\usepackage[T1]{fontenc}
\usepackage[utf8]{inputenc}
\usepackage[english]{babel}
\usepackage{geometry}
\usepackage{standalone}
\usepackage{graphicx}
\usepackage[dvipsnames,svgnames,x11names]{xcolor}
\usepackage{mathtools,bm,amssymb}
\usepackage{enumitem}
\usepackage{booktabs}
\usepackage{natbib}
\usepackage[strict=true]{csquotes}
\graphicspath{{./figs/}{.}}
\usepackage{setspace}
\usepackage{Sweave}
\usepackage{alltt}
\usepackage{framed}
\usepackage{marginnote}

\newcommand{\E}{\mathbb{E}}

\newcommand{\Real}{\ensuremath{\mathbb{R}}}
\newcommand{\R}{\Real}
\newcommand{\CondInd}[3]{#1 \perp\kern-5pt \perp #2 \mid #3}
\newcommand{\Ind}[2]{#1 \perp\kern-5pt \perp #2}

\DeclareMathOperator{\diag}{diag}

\newcommand{\independenT}[2]{\mathrel{\setbox0\hbox{$#1#2$}\copy0\kern-\wd0\mkern4mu\box0}}

\newcommand{\minp}{\operatorname{MIN}}

\title{A general approach to construct powerful tests for intersections of one-sided null-hypotheses based on influence functions}
\author{Christian Bressen Pipper${}^{1,2}$,
    Andreas Nordland${}^1$,\\
    Klaus Kähler Holst${}^{1}$}
  \date{${}^1$Novo Nordisk, Søborg, Denmark \\
    ${}^2$Department of Public Health, Epidemiology, Biostatistics and
    Biodemography, University of Southern Denmark, Odense, Denmark
    \\[2ex]%
    \today}

\begin{document}
    
\maketitle


\begin{abstract}
\noindent  
Testing intersections of null-hypotheses is an integral part of closed testing procedures for assessing multiple null-hypotheses under family-wise type 1 error control. Popular intersection tests such as the minimum p-value test are based on marginal p-values and are typically evaluated conservatively by disregarding simultaneous behavior of the marginal p-values.  
We consider a general purpose Wald type test for testing intersections of one-sided null-hypotheses. The test is constructed on the basis of the simultaneous asymptotic behavior of the p values. The simultaneous asymptotic behavior is derived via influence functions of estimators using the so-called stacking approach. In particular, this approach does not require added assumptions on simultaneous behavior to be valid. The resulting test is shown to have attractive power properties and thus forms the basis of a powerful closed testing procedure for testing multiple one-sided hypotheses under family-wise type 1 error control. 
\end{abstract}


\section{Introduction}\label{sec:intro}

The vast majority of clinical trials aim to answer multiple research questions in a scientifically reliable manner. This is particularly true for pivotal clinical trials in drug development where, for instance, treatment efficacy is often assessed in several clinical endpoints, reflecting different clinically important aspects of potential treatment benefit. From a statistical perspective, this requires that multiple one-sided null-hypotheses are tested under family-wise type 1 error control, that is, controlling the probability of erroneously rejecting one or more true null-hypotheses. Family-wise type 1 error control is, in turn, ensured by applying an appropriate multiple testing procedure \cite[Chapter~2]{Dmitrienko2010}.

Popular choices of multiple testing procedures include fixed sequence testing, fall back procedures, and more generally sequentially rejective multiple testing procedures \citep{bretz2009}. The common denominator of all these proposals is the closure principle that ensures family-wise type 1 error control \citep{closedtesting}. 

To reject a specific null-hypothesis according to the closure principle, one needs to reject pre-specified valid $\alpha$-level tests for the intersection of each subset of considered null-hypotheses when that subset includes the specific null-hypothesis. Here a valid $\alpha$-level test for a specific null-hypothesis refers to a test, where the probability of rejecting this null-hypothesis does not exceed $\alpha$ if the null-hypothesis is true.

It is clear from the closure principle that its ability to reject a null-hypothesis relies critically on the ability to construct powerful valid $\alpha$-level tests for all relevant intersections of null-hypotheses. Current popular constructions include the minimum weighted p-value intersection tests that underpin sequentially rejective weighted Bonferroni methods \citep{hommel2007,bretz2009,Bretz2011}. These intersection tests do not rely on the simultaneous behavior of the marginal p-values for the separate null-hypotheses, and it has been shown that power-gains can be achieved without sacrificing $\alpha$-level validity by utilizing the simultaneous behavior of the marginal p-values \citep{Bretz2011}.

As discussed in \cite{Bretz2011} this potential improvement requires a specification of the
joint distribution of the marginal p-values, which has traditionally been done
by imposing further assumptions about joint behavior. Specifically, joint behavior is most often derived from a parametric model encompassing the unknown treatment effects that are targeted by the marginal hypotheses. In this case it is well known that likelihood ratio type tests for the intersection hypothesis will have superior power performance in terms of for instance Bahadur efficiency \citep[Chapter~16]{vaart_1998_asymp}.

In this paper we consider a Wald type variant of the likelihood ratio test for testing the intersection of one-sided hypotheses as proposed in \cite{silvapulle92_robustwald}. Our proposal utilizes the joint asymptotic normality of the estimated treatment effects that are assessed in the one-sided null-hypotheses, both when calculating the test-statistic, where the variance/covariance matrix is used actively, but also when characterizing the asymptotic distribution of the test statistic in order to calculate p-values.

The joint asymptotic normality, and in particular an estimator of the variance/covariance matrix, is obtained via the stacking approach \citep{Pipper2012}, that is, by appealing to the asymptotic linearity of each single estimated treatment effect but without imposing any additional assumptions about joint behavior. 

In the special case where the intersection null hypothesis is the intersection of two one-sided null-hypotheses we derive closed form expressions of the asymptotic distribution of the test statistic under the null-hypothesis. In this special case we also provide a theoretical comparison with the minimum p-value test based on the joint distribution of the marginal p-values derived from the stacking approach. We show that our proposal is indeed superior in terms of Bahadur efficiency. 

For the general case where the intersection null hypothesis is the intersection of more than two one-sided null-hypotheses we also provide a characterization of the asymptotic null distribution that allows us to compute a p-value via simulation. An additional challenge in this setting is that the test statistic has no closed form expression and needs to be calculated via an optimization algorithm due to \cite{Dykstra1983, Dykstra1986}.

We also provide a weighted version of the proposed test and derive its asymptotic properties. The weighted version of the test allows users to reflect the relative importance among the separate one-sided hypotheses by assigning higher weights to more important hypotheses. Conceptually this is similar to the sequentially rejective weighted Bonferroni methods \citep{hommel2007,bretz2009,Bretz2011}, but instead of weighing the marginal p-values, we propose weighing the estimated treatment effects. 

Finally, we describe a software implementation through the R-package targeted and assess the performance of the proposed methodology in the context the FLOW trial \cite{Flow2024}, which is a recently conducted large scale randomized controlled trial. 

The paper is structured as follows. We introduce the formal setup and notation in Section \ref{sec:setup}. Section \ref{sec:testing} is dedicated to large sample properties in the special case where the intersection null hypothesis is the intersection of two one-sided null-hypotheses. Section \ref{sec:asympgen} considers large sample properties in the general case. In section \ref{sec:sims} we investigate small sample performance of the proposed methodology under several scenarios including scenarios mimicking the FLOW trial, and we then proceed to analyze the FLOW trial in Section \ref{sec:flowana}. A discussion is provided in Section \ref{sec:disc}.

\section{Setup and notation}\label{sec:setup}

In our setup we consider target treatment effects $\theta_{j}, j=1,\ldots,J$ with corresponding estimators $\hat{\theta}_{j}$ based on independent identically distributed subject specific data $Z_{i}$ from a sample of $n$ individuals. We assume that the estimators are asymptotically linear with influence function $\phi_{j}$. Specifically, we assume that
\begin{equation}\label{LAN}
\sqrt{n}\cdot(\hat{\theta}_{j}-\theta_{j})=\frac{1}{\sqrt{n}}\sum_{i=1}^{n}\phi_{j}(Z_{i})+o_{P}(1).
\end{equation}

As a consequence we may stack the influence functions \citep{Pipper2012} to obtain joint asymptotic normality, that is, with $\hat{\theta}=(\hat{\theta}_{1},\ldots,\hat{\theta}_{J})$ and $\theta=(\theta_{1},\ldots,\theta_{J})$ we have the following weak convergence result:

\begin{equation}\label{jointMVN}
\sqrt{n}\cdot(\hat{\theta}-\theta)\rightsquigarrow \mathcal{N}(0,\Sigma),
\end{equation}
where
$\Sigma=\E\{\phi(Z_{i})^{\otimes 2}\}$
with $\phi(Z_{i})=\{\phi_{1}(Z_{i}),\ldots,\phi_{J}(Z_{i})\}$.

As noted in \cite{Pipper2012} a consistent estimator of $\Sigma$ can be obtained as:

$$
\hat{\Sigma}=\frac{1}{n}\sum_{i=1}^{n}\hat{\phi}(Z_{i})^{\otimes 2},
$$
where $\hat{\phi}(Z_{i})$ denotes the empirical counterpart of $\phi(Z_{i})$.

To assess individual treatment effects quantified by $\theta_{j}$ we now consider the following one-sided null-hypotheses 

$$H_{j}: \theta_{j}\leq \delta_{j}, j=1,\ldots,J $$
with $\delta_{j}\in\R$ denoting non-inferiority/superiority margins. 

We propose to test the intersection hypothesis $\cap_{j=1}^{J}H_{j}$ at $\alpha$ level by means of a Wald test proposed in for instance \citep[p. 224]{robertson1988order} or \citep{silvapulle92_robustwald}. In our particular context, we consider a version of this test that is truncated at zero for values below zero, and we term this the signed Wald test in what follows. Accordingly, the signed Wald test is defined as follows:
\begin{equation}\label{SW}
SW_{n,\cap_{j=1}^{J} H_{j}}=\inf_{\theta\in \cap_{j=1}^{J}H_{j}} \big\{n\cdot\{\hat{\theta}-\theta\}^{\top}\hat{\Sigma}^{-1}\{\hat{\theta}-\theta\}\big\}.
\end{equation}

Another popular choice of test statistic for multi-purpose testing of intersections of null-hypotheses is the min p-value test \citep{hommel2007} given by:

\begin{equation}\label{minp}
\minp(p)_{n,\cap_{j=1}^{J} H_{j}}=\min_{j=1,\ldots,J}\{p_{n,j}\}.
\end{equation}
Here $p_{n,j}$  denote  the single hypothesis p-values, that is: 

$$
p_{n,j}=1-\Phi(\sqrt{n}\cdot\frac{\hat{\theta}_{j}-\delta_{j}}{\sqrt{\hat{\Sigma}_{jj}}}),
$$
where $\Phi$ denotes the cumulative distribution function of the standard normal distribution. We note that the min p-value test can  be reformulated as

$$
\minp(p)_{n,\cap_{j=1}^{J} H_{j}}=1-\Phi(\sqrt{n}\cdot Z_{n,max})
$$
with
$$
Z_{n,max}=\max_{j=1,\ldots, J}\big\{\frac{\hat{\theta}_{j}-\delta_{j}}{\sqrt{\hat{\Sigma}_{jj}}}\}.
$$
This shows that using the min p-value test statistic is equivalent to using the test statistic $\sqrt{n}\cdot Z_{n,max}$.

In the above signed Wald test we don't enforce any priority among the single hypotheses $H_{j}$. However, this can be enforced by introducing weights $w_{j}\geq0$ with $\sum_{j=1}^{J}w_{j}=1$ and considering the weighted signed Wald test given by:

\begin{equation}\label{SWweighted}
SW_{n,\cap_{j=1}^{J} H_{j},w}=\inf_{\theta\in \cap_{j=1}^{J}H_{J}} \big\{n\cdot\{\hat{\theta}-\theta\}^{\top}\cdot W\cdot \hat{\Sigma}^{-1}\cdot W \cdot \{\hat{\theta}-\theta\}\big\}
\end{equation}
with
$$
W=\diag\{w_{1},\ldots,w_{J}\}.
$$

Note that (\ref{SWweighted}) is indeed a generalization of (\ref{SW}) since $SW_{n,\cap_{j=1}^{J} H_{j}}=J^{2}\cdot SW_{n,\cap_{j=1}^{J} H_{j},w}$ with $w_{1}=\ldots=w_{J}=\frac{1}{J}$. For the remainder of the paper we shall consider the signed Wald test as a special case of the weighted signed Wald test with equal weights. 

Also note that compared to the weighted min p-value test \citep{Bretz2011}, the above weighting proposal operates on the quantification scale rather than the p-value scale. As a consequence, the two approaches may perform quite differently with the same weights making them hard to compare in general. However, for the special case where no priority among hypotheses is induced, a direct comparison is both meaningful and instructive. We therefore provide a theoretical comparison between the signed Wald test and the min p-value test in Section \ref{sec:bahadur} as well as a comparison of power performance through simulation in Section \ref{sec:sims}.

\section{Large sample properties when testing the intersection of two one-sided hypotheses}\label{sec:testing}

For the two hypothesis case we derive a closed form expression of the weighted signed Wald test leading to an easily applicable asymptotic approximation of the p-value. To motivate the use of the weighted signed Wald test from a power perspective, we also provide a theoretical comparison in the special case of the signed Wald test and the minimum p-value test in terms of Bahadur efficiency.  

\subsection{Asymptotic properties}\label{asymp:two}
In order to derive large sample properties of  $SW_{n,H_{1}\cap H_{2}},w$ we first rewrite above expression in terms of $\hat{u}=\sqrt{n}\cdot\sqrt{\hat{\Sigma}^{-1}}\cdot W\cdot \{\hat{\theta}-\delta\}$ and $u=\sqrt{n}\cdot\sqrt{\hat{\Sigma}^{-1}}\cdot W\cdot\{\theta-\delta\}$ with $\delta=(\delta_{1},\delta_{2})$ to obtain:
\begin{equation}\label{wSignWald}
SW_{n,H_{1}\cap H_{2},w}=\inf_{W^{-1}\cdot\sqrt{\hat{\Sigma}}\cdot u\leq 0} \big\{\{\hat{u}-u\}^{\top}\{\hat{u}-u\}\big\}=\inf_{\sqrt{\hat{\Sigma}}\cdot u\leq 0} \|\hat{u}-u\|^{2},
\end{equation}

\begin{figure}[htpb]
  \centering
  \includegraphics[width=0.7\textwidth]{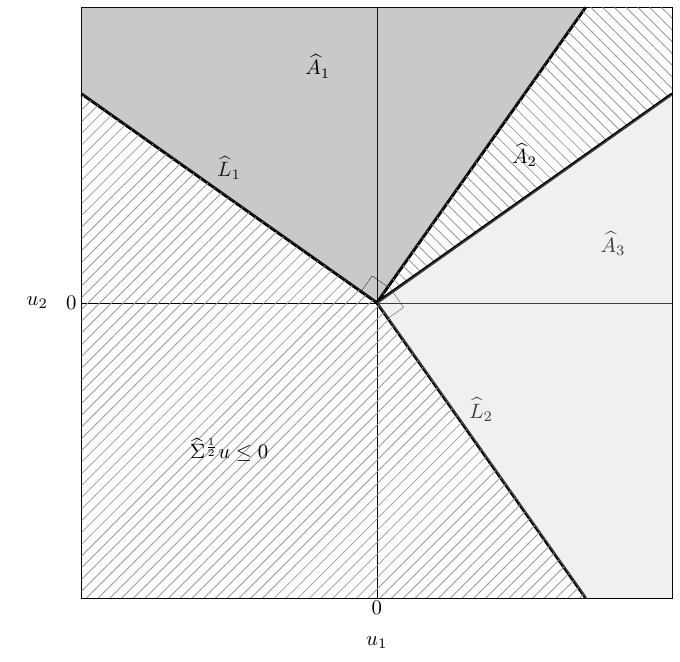}
  \caption{Regions characterizing the value of the signed Wald test}
  \label{fig:wald}
\end{figure}

As illustrated in Figure \ref{fig:wald} the region $\{u: \sqrt{\hat{\Sigma}}\cdot u\leq0\}$ is enclosed by the two lines $\hat{L}_{1}$ and $\hat{L}_{2}$. Note that if $\hat{u}$ belongs to that  region the weighted signed wald test equals zero. If $\hat{u}\in\hat{A}_{1}$ we know that the projection of $\hat{u}$ onto $\hat{L}_{1}$ is the point in  $\{u: \sqrt{\hat{\Sigma}}\cdot u\leq0\}$ closest to $\hat{u}$. Accordingly, for $\hat{u}\in\hat{A}_{1}$, we have $SW_{n,H_{1}\cap H_{2},w}=\|\hat{u}-P_{\hat{L}_{1}}(\hat{u})\|^{2}$, where $P_{\hat{L}_{1}}(\hat{u})$ denotes the projection of $\hat{u}$ onto $\hat{L}_{1}$. Similarly, it follows that $SW_{n,H_{1}\cap H_{2},w}=\|\hat{u}-P_{\hat{L}_{2}}(\hat{u})\|^{2}$ for $\hat{u}\in\hat{A}_{3}$. Finally, for $\hat{u}\in\hat{A}_{2}$ the point in  $\{u: \sqrt{\hat{\Sigma}}\cdot u\leq0\}$ closest to $\hat{u}$ is zero and accordingly $SW_{n,H_{1}\cap H_{2},w}=\|\hat{u}\|^{2}$ in this case.

In summary, we conclude that the weighted signed Wald test for $H_{1}\cap H_{2}$ may be rewritten as:  
$$
SW_{n,H_{1}\cap H_{2},w}=I(\hat{u}\in\hat{A}_{1})\cdot\|\hat{u}-P_{\hat{L}_{1}}(\hat{u})\|^{2}+ I(\hat{u}\in\hat{A}_{3})\cdot\|\hat{u}-P_{\hat{L}_{2}}(\hat{u})\|^{2}+
I(\hat{u}\in\hat{A}_{2})\cdot\|\hat{u}\|^{2}
$$

Next note that when $\theta=\delta$ we have that $\hat{u}$ converges weakly to a zero mean normal distribution with variance $\sqrt{\Sigma^{-1}}\cdot W \cdot\Sigma\cdot W\cdot\sqrt{\Sigma^{-1}}$,  where $\Sigma$ denotes the positive definite limit in probability of $\hat{\Sigma}$.  It follows that: 

\begin{align*}
&I(\hat{u}\in\hat{A}_{1})\cdot\|\hat{u}-P_{\hat{L}_{1}}(\hat{u})\|^{2}\overset{D}\rightarrow I(U\in A_{1})\frac{(\beta_{1}\cdot U_{1}-U_{2})^{2}}{1+\beta_{1}^{2}},\\
&I(\hat{u}\in\hat{A}_{3})\cdot\|\hat{u}-P_{\hat{L}_{2}}(\hat{u})\|^{2}
\overset{D}\rightarrow I(U\in A_{3})\frac{(\beta_{2}\cdot U_{2}-U_{1})^{2}}{1+\beta_{2}^{2}},\\
&I(\hat{u}\in\hat{A}_{2})\cdot\|\hat{u}\|^{2}\overset{D}\rightarrow I(U\in A_{2})\cdot(U_{1}^{2}+U_{2}^{2}),
\end{align*}
where $U=(U_{1},U_{2})\sim \mathcal{N}(0,\sqrt{\Sigma^{-1}}\cdot W \cdot \Sigma\cdot W\cdot \sqrt{\Sigma}^{-1})$, $A_{j},\: j=1,2,3$ corresponds to $\hat{A}_{j}, \:j=1,2,3$ when replacing $\hat{\Sigma}$ with $\Sigma$, and  $\beta_{j},\:j=1,2$ is the limiting slope of $\hat{L}_{j}$.

It follows that the p-value, that is, the maximal tail probability in the distribution of $SW_{n,H_{1}\cap H_{2},w}$ under the null hypothesis, can be approximated as 
\begin{eqnarray*}
  &\sup_{\theta\in H_{1}\cap H_{2}}P_{\theta}(SW_{n,H_{1}\cap H_{2},w}\geq x)&=P_{\theta=\delta}(SW_{n,H_{1}\cap H_{2},w}\geq x) \\
  & &\quad \longrightarrow P(SW_{H_{1}\cap H_{2},w}\geq x), as\:\: n\rightarrow\infty
\end{eqnarray*}
where
\begin{equation}\label{wmixdist}
SW_{H_{1}\cap H_{2},w}=I(U\in A_{1})\frac{(\beta_{1}\cdot U_{1}-U_{2})^{2}}{1+\beta_{1}^{2}}+I(U\in A_{3})\frac{(\beta_{2}\cdot U_{2}-U_{1})^{2}}{1+\beta_{2}^{2}}+I(U\in A_{2})\cdot(U_{1}^{2}+U_{2}^{2}).
\end{equation}
In practice, the p value can be calculated through simulating $SW_{H_{1}\cap H_{2},w}$, where we have replaced $\Sigma$ by $\hat{\Sigma}$ in all relevant quantities.

Also note that in the special case $w_{1}=w_{2}=\frac{1}{2}$ ,corresponding to the signed Wald test (\ref{SW}), we have 

$$\sqrt{\Sigma^{-1}}\cdot W \cdot \Sigma\cdot W\cdot \sqrt{\Sigma}^{-1}=\sqrt{\hat{\Sigma}^{-1}}\cdot W \cdot \hat{\Sigma}\cdot W\cdot \sqrt{\hat{\Sigma}}^{-1}=W^{2}=\frac{1}{4}I_{2\times 2}.$$
As a consequence the above approximation of the p-value simplifies substantially due to the fact that $P(U\in A_{1})=P(U\in A_{3})=\frac{1}{4}$ and since in this case 

\begin{align*}
&4\cdot \frac{(\beta_{1}\cdot U_{1}-U_{2})^{2}}{1+\beta_{1}^{2}}\sim \chi^{2}_{1},\\
&4\cdot \frac{(\beta_{2}\cdot U_{2}-U_{2})^{1}}{1+\beta_{2}^{2}}\sim \chi^{2}_{1},\\
&4\cdot(U_{1}^{2}+U_{2}^{2})\sim\chi^{2}_{2}.
\end{align*}
For more details see \cite{holst2025}. 

\subsection{Bahadur relative efficiency}\label{sec:bahadur}
Bahadur relative efficiency is a mathematically tractable way to compare the large sample performance of two test statistics in terms of sample size required to achieve a certain power \citep{bahadur1960}. Specifically, for two sequences of $\alpha$ level test statistics $T_{j,n}, \: j=1,2$, let $N_{j}(\alpha,\beta,\theta^{\ast})$ denote the smallest sample size required to achieve power $\beta$ under a given alternative $\theta^{\ast}$. Then the Bahadur relative efficiency in that alternative is defined as the following limit:

$$
BRE_{T_{1,n},T_{2,n}}(\theta^{\ast})=\lim_{\alpha\rightarrow0}\frac{N_{1}(\alpha,\beta,\theta^{\ast})}{N_{2}(\alpha,\beta,\theta^{\ast})}
$$.

Accordingly, the Bahadur relative efficiency approximates the relative gain in required sample size, at least when tests are performed at a low $\alpha$ level. 

The Bahadur relative efficiency can be further characterized as the ratio of Bahadur slopes \citep{Rogers1988}, where the Bahadur slope of $T_{j,n}$ with accompanying p-value $p_{j,n}$ is defined as:
$$
d_{j}(\theta^{\ast})=-2\lim_{n\rightarrow\infty}\frac{\log(p_{j,n})}{n}.
$$

As a first step towards deducing the Bahadur slope of our proposed intersection test we characterize the limiting value of $n^{-1}SW_{n,H_{1}\cap H_{2}}$. To this end note that by brute force calculation we may show that 

\begin{eqnarray}\label{sw:repr}
 &&SW_{n,H_{1}\cap H_{2}}=n\cdot\big\{I\big(Z_{n,max}\geq 0,\: Z_{n,min}\leq\hat{\rho}\cdot Z_{n,max}\big)\cdot Z_{n,max}^{2}+\\
 &&+I\big(Z_{n,max}\geq 0,\: Z_{n,min}\geq\hat{\rho}\cdot Z_{n,max}\big)\frac{(Z_{n,max}-Z_{n,min})^{2}+2\cdot(1-\hat{\rho})\cdot Z_{n,min}\cdot Z_{n,max}}{1-\hat{\rho}^{2}}\big\}\nonumber
\end{eqnarray}
 with
\begin{align*}
&Z_{n,max}=\max\{\frac{\hat{\theta}_{1}-\delta_{1}}{\sqrt{\hat{\Sigma}_{11}}},\frac{\hat{\theta}_{2}-\delta_{2}}{\sqrt{\hat{\Sigma}_{22}}}\},\\
&Z_{n,min}=\min\{\frac{\hat{\theta}_{1}-\delta_{1}}{\sqrt{\hat{\Sigma}_{11}}},\frac{\hat{\theta}_{2}-\delta_{2}}{\sqrt{\hat{\Sigma}_{22}}}\},\\
&\hat{\rho}=\frac{\hat{\Sigma}_{12}}{\sqrt{\hat{\Sigma}_{11}\cdot\hat{\Sigma}_{22}}}. 
\end{align*}
From the above characterization we note that 

\begin{align}\label{SWnoncent}
\begin{split}
n^{-1}SW_{n,H_{1}\cap H_{2}}\overset{P}{\rightarrow}& I\big(z_{max}\geq 0,\: z_{min}\leq\rho\cdot z_{,max}\big)\cdot z_{max}^{2}\\
 &+I\big(z_{max}\geq 0,\: z_{,min}\geq\rho\cdot z_{max}\big)\frac{(z_{max}-z_{min})^{2}+2\cdot(1-\rho)\cdot z_{min}\cdot z_{max}}{1-\rho^{2}} 
\end{split}
\end{align}
where 

\begin{align*}
&z_{max}=\max\{\frac{\theta_{1}-\delta_{1}}{\sqrt{\Sigma_{11}}},\frac{\theta_{2}-\delta_{2}}{\sqrt{\Sigma_{22}}}\},\\
&z_{min}=\min\{\frac{\theta_{1}-\delta_{1}}{\sqrt{\Sigma_{11}}},\frac{\theta_{2}-\delta_{2}}{\sqrt{\Sigma_{22}}}\},\\
&\rho=\frac{\Sigma_{12}}{\sqrt{\Sigma_{11}\cdot\Sigma_{22}}}. 
\end{align*}

It now follows from Lemma 3 in \cite{Rogers1988} and the characterization of the Bahadur slope in that paper that the Bahadur slope of $SW_{n,H_{1}\cap H_{2}}$ is given by (\ref{SWnoncent}).

We next turn to the Bahadur slope for the min p-value test. Since the min p-value test is equivalent to a one-sided test based on the test statistic $\sqrt{n}\cdot Z_{n,max}$ (see Section \ref{sec:setup}) we proceed by calculating the Bahadur slope of this test and note that due to the equivalence this slope equals that of the min p-value test.  If we again use the characterization of the Bahadur slope in \cite{Rogers1988} and repeat the steps of Lemma 1 in that paper, we conclude that the Bahadur slope of the min p-value test is given by:

\begin{equation}\label{minp-slope}
z_{max}^{2}.
\end{equation}

Finally, we note that the ratio between (\ref{SWnoncent}) and (\ref{minp-slope}) under the alternative ($z_{max}>0$) is 1 when $z_{min}\leq\rho\cdot z_{max}$ and increases from 1 to $\frac{2}{1+\rho}>1$ for  $\rho\cdot z_{max}<z_{min}\leq z_{max}$. We conclude that when testing the intersection of two one-sided hypotheses the signed Wald test performs better than the min p-value test in terms of Bahadur efficiency.

\section{Large sample properties in the general case}\label{sec:asympgen}

For the general case we also define $u=\sqrt{n}\cdot\sqrt{\hat{\Sigma}^{-1}}\cdot W\cdot(\theta-\delta)$ and $\hat{u}=\sqrt{n}\cdot\sqrt{\hat{\Sigma}^{-1}}\cdot W\cdot(\hat{\theta}-\delta)$ and, as in the two hypothesis case,  proceed to rewrite the weighted signed Wald test as:

\begin{equation}\label{WSW:repr}
SW_{n,\cap_{j=1}^{J} H_{j},w}=\inf_{\sqrt{\hat{\Sigma}}\cdot u\leq 0} \|\hat{u}-u\|^{2}.
\end{equation}
It now follows from standard asymptotic arguments that when $\theta=\delta$ we have: 

\begin{equation}\label{limitdistweight}
SW_{n,\cap_{j=1}^{J} H_{j},w}\overset{D}\rightarrow \inf_{\{u\in\mathbf{R}^{J}|\:\sqrt{\Sigma}\cdot u\leq 0\}} \|U-u\|^{2},
\end{equation}
where $U\sim \mathcal{N}(0,\sqrt{\Sigma^{-1}}\cdot W \cdot \Sigma\cdot W\cdot \sqrt{\Sigma}^{-1})$.

In this general case there is no readily available expression \citep{Dykstra1983, Dykstra1986} that allows us to directly calculate  $SW_{n,\cap_{j=1}^{J} H_{j},w}$ in terms of $\hat{u}$ and $\hat{\Sigma}$. Similarly, we are unable to directly calculate the limiting distributions in terms of $U$ and $\Sigma$. Instead, we use a numerical procedure owing to \cite{Dykstra1983} to approximate  $SW_{n,\cap_{j=1}^{J} H_{j},w}$. We also use this procedure to approximate the limiting distribution as follows. We generate a large number of realizations of $U$ according to the normal distribution $\mathcal{N}(0,\sqrt{\hat{\Sigma}^{-1}}\cdot W \cdot \hat{\Sigma}\cdot W\cdot \sqrt{\hat{\Sigma}^{-1}})$. Subsequently, we plug in each of these values and  $\hat{\Sigma}$ and approximate the right-hand side of  (\ref{limitdistweight})  using the algorithm outlined in \cite{Dykstra1983}. The empirical distribution of the resulting approximate realizations of the right-hand side of (\ref{limitdistweight}) then forms the basis for calculating the p-value as the fraction of realizations that exceed the calculated value of the test statistic. For completeness, we give a short description of the approximation algorithm below.

First we note that 

$$
\{u\in \mathbf{R}^{J}: \:\sqrt{\hat{\Sigma}}\cdot u\leq 0\}=\cap_{j=1}^{J}K_{j},
$$
where 
$$
K_{j}=\{u\in\mathbf{R}^{J}:\: \sum_{l=1}^{J}\sqrt{\hat{\Sigma}}_{jl}\cdot u_l\leq 0\}.
$$

This makes Dykstras projection algorithm particularly attractive due to the fact that it iterates through projections on each of the half planes $K_{j}$. These projections in turn are very easy to calculate according to well known closed form expressions. 

Specifically in the context of (\ref{WSW:repr}), Dykstras projection algorithm \citep{Dykstra1983} produces a sequence of vectors $\hat{u}_{n,j}\rightarrow u^{\ast}$ as $n\rightarrow\infty$ for any $j=1,\ldots, J$ such that

$$
SW_{n,\cap_{j=1}^{J} H_{j},w}=\|\hat{u}-u^{\ast}\|.
$$
The vector $\hat{u}_{n,j}$ is determined recursively as the projection onto $K_{j}$ of 

$$
\hat{u}+\sum_{l=1}^{j-1}\Delta_{n,l}+\sum_{l=j+1}^{J}\Delta_{n-1,l}
$$
followed by calculating the increment
$$
\Delta_{n,j}=\hat{u}_{n,j}-(\hat{u}+\sum_{l=1}^{j-1}\Delta_{n,l}+\sum_{l=j+1}^{J}\Delta_{n-1,l}).
$$

\section{Simulation study}\label{sec:sims}

To assess the numerical performance of the proposed intersection tests we conduct three simulation studies. 
The intersection test is implemented in full generality in the targeted R package \citep{targetedr} and
implementation details in the context of simulation study 3 are given in Appendix \ref{sec:software}.

\begin{figure}[htpb]
  \centering
  \includegraphics[width=0.7\textwidth]{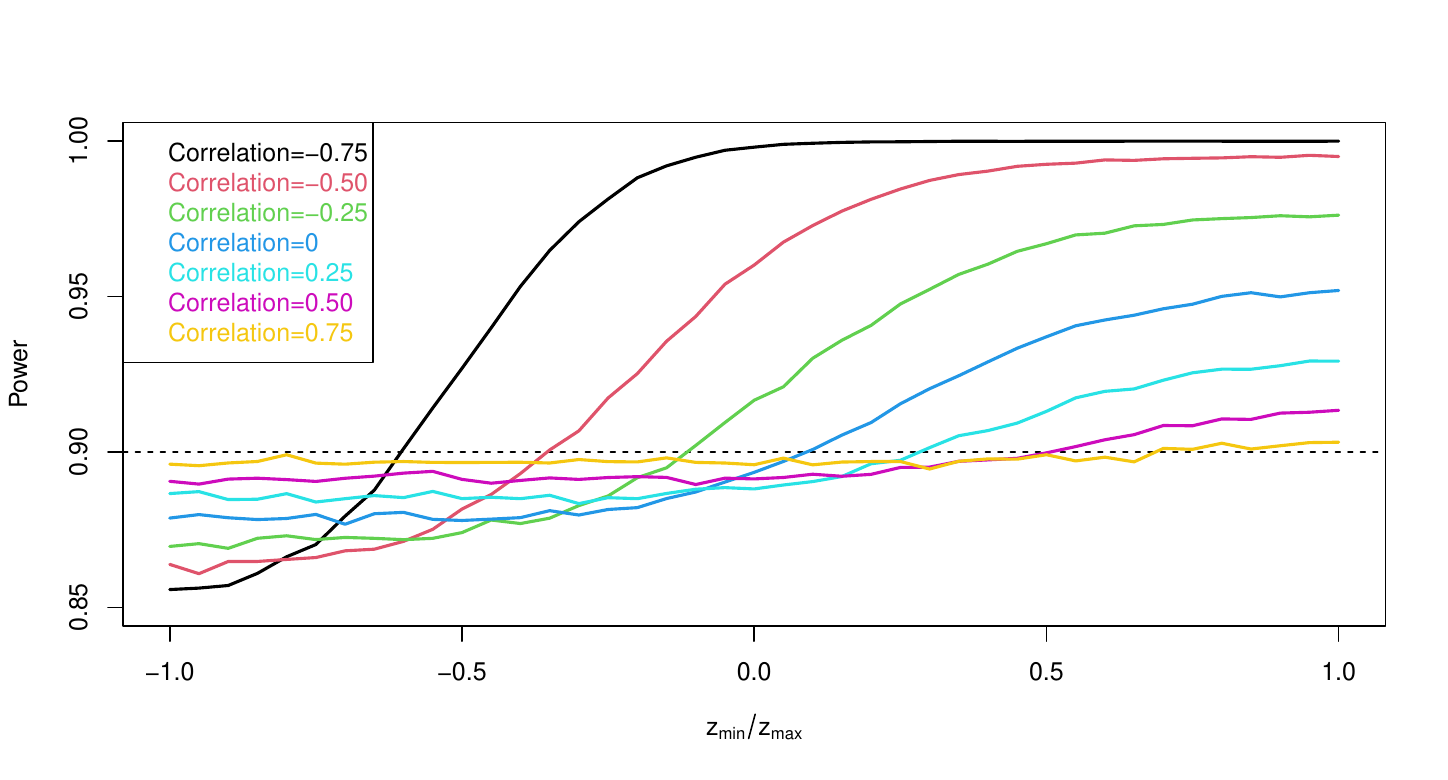}
  \caption{Power of the signed Wald test as a function of $z_{min}/z_{max}$ for a range of correlations. The Dashed line corresponds to power of the minimal p-value test}
  \label{fig:powertwo}
\end{figure}

 In the first simulation study, we consider the two hypothesis scenario with $\delta=0$ and compare the signed Wald test to the min p-value test in terms of power. Specifically, we consider the range of correlations $\rho=-0.75,-0.5,-0.25,0,0.25,0.5,0.75$. For each correlation we consider alternatives $\Big(\max\{\frac{\theta_{1}}{\sqrt{\Sigma_{11}}},\frac{\theta_{2}}{\sqrt{\Sigma_{22}}}\},\min\{\frac{\theta_{1}}{\sqrt{\Sigma_{11}}},\frac{\theta_{2}}{\sqrt{\Sigma_{22}}}\}\Big)=(z_{max}, z_{min})=(z_{max},s\cdot z_{max}), \:s\in \{-1,-0.95,\ldots,0.95,1\} $ and with a sample size chosen so that the minimal p-value test $Z_{n,max}$ yields a power of $0.9$ with a significance level $\alpha=0.025$. For each correlation $\rho$ and each alternative $(z_{max},s\cdot z_{min})$ we generate 100,000 independent realizations

 $$
 \Big(\frac{\hat{\theta}_{1}}{\sqrt{\hat{\Sigma}_{11}}},\frac{\hat{\theta}_{2}}{\sqrt{\hat{\Sigma}_{22}}}\Big)\sim N\Big((z_{max},z_{min}), \begin{pmatrix} 1 & \rho\\ \rho & 1 \end{pmatrix}\Big).
 $$
For each realization we calculate the signed Wald test using (\ref{sw:repr}) and reject the intersection of null hypotheses if the value falls below the critical value corresponding to the correlation $\rho$ and significance level $\alpha=0.025$ (see Section \ref{asymp:two} or \cite{holst2025} for details on calculating the critical value).  The simulated power is calculated as the fraction of rejections and are presented in Figure \ref{fig:powertwo}.

From Figure \ref{fig:powertwo} we see substantial power gains when both null-hypotheses are false and correlation is moderate or negative. We also see that the power gains increase as correlation decreases. Our findings are well aligned with the theoretical results on Bahadur relative efficiency derived in Section \ref{sec:bahadur}. Here we also found a favorable as well as increasing relative efficiency with decreasing correlation.  

In our second simulation study we investigate the small sample performance in terms of type 1 error control. Our simulation setup is tailored to mirror the FLOW study we analyse later in this paper. Specifically, we consider three null hypotheses and intersections thereof as specified below.

We let $A$ denote a dichotomous treatment indicator corresponding to a 1:1 randomized treatment intervention. We let $R$ denote the occurrence of a terminal event before some landmark time. Finally, we let $Y$ denote a score value also recorded at the same landmark time but only in the absence of the terminal event. If the terminal event occurs before landmark we artificially assign the score to some unfavorable value $\Gamma$ leading to the composite score $\tilde{Y}=(1-R)\cdot Y+R\cdot \Gamma$, which then reflects score progression  balanced by a penalty/price due to occurrence of terminal event. In this setup we consider the following three treatment contrasts:

\begin{align*}
&\theta_{1}=P(R=0|A=1)-P(R=0| A=0),\\
&\theta_{2}=E(Y|R=0,A=1)-E(Y|R=0,A=0),\\
&\theta_{3}=E(\tilde{Y}|A=1)-E(\tilde{Y}|A=0),
\end{align*}
and assess the intersections of subsets of the corresponding three superiority hypotheses $H_{j}:\: \theta_{j}\leq 0$. 

The assessment will be based on the one-step estimators outlined in \cite{holst2025} which in the the above scenario with no baseline covariates and no missing endpoint values reduce to:

\begin{align*}
    &\hat{\theta}_{1}=\frac{\sum_{i=1}^{n}A_{i}\cdot (1-R_{i})}{\sum_{i=1}^{n}A_{i}}-\frac{\sum_{i=1}^{n}(1-A_{i})\cdot (1-R_{i})}{\sum_{i=1}^{n}(1-A_{i})},\\
    &\hat{\theta}_{2}=\frac{\sum_{i=1}^{n}A_{i}\cdot Y_{i}\cdot (1-R_{i})}{\sum_{i=1}^{n}(1-R_{i})\cdot A_{i}}-\frac{\sum_{i=1}^{n}(1-A_{i})\cdot Y_{i}\cdot (1-R_{i})}{\sum_{i=1}^{n}(1-R_{i})\cdot(1-A_{i})},\\
    &\hat{\theta}_{3}=\frac{\sum_{i=1}^{n}A_{i}\cdot\tilde{Y}_{i}}{\sum_{i=1}^{n}A_{i}}-\frac{\sum_{i=1}^{n}(1-A_{i})\cdot\tilde{Y}_{i}}{\sum_{i=1}^{n}(1-A_{i})},\\
\end{align*}
where $Z_{i}=(A_{i},R_{i},Y_{i})$ are independent realizations of $Z=(A,R,Y)$ and $\tilde{Y}_{i}=(1-R_{i})\cdot Y_{i}+R_{i}\cdot\Gamma$.

The corresponding influence functions were also derived in \cite{holst2025}. For completeness, we note that in our simplified scenario they are given by: 

\begin{align*}
   &\phi_{1}(Z_{i})=\psi_{11}(Z_{i})-\psi_{10}(Z_{i}),\\
   &\phi_{2}(Z_{i})=\psi_{21}(Z_{i})-\psi_{20}(Z_{i}),\\
   &\phi_{3}(Z_{i})=\psi_{31}(Z_{i})-\psi_{30}(Z_{i}),
\end{align*}

\begin{align*}
    \psi_{10}(Z_{i})=&\frac{(1-R_{i})\cdot(1- A_{i})-P(R=0,A=0)}{P(A=0)}-\frac{P(R=0,A=0)}{P(A=0)^{2}}\cdot\{1-A_{i}-P(A=0)\},\\
    \psi_{11}(Z_{i})=&\frac{(1-R_{i})\cdot A_{i}-P(R=0,A=1)}{P(A=1)}-\frac{P(R=0,A=1)}{P(A=1)^{2}}\cdot\{A_{i}-P(A=1)\},\\
    \psi_{20}(Z_{i})=&\frac{Y_{i}\cdot (1-R_{i})\cdot(1-A_{i})-E\{Y\cdot(1-R)\cdot(1-A)\}}{P(R=0,A=0)}\\
    &-\frac{E\{Y\cdot(1-R)\cdot(1-A)\}}{P(R=0,A=0)^{2}}\{(1-R_{i})\cdot (1-A_{i})-P(R=0,A=0)\},\\
    \psi_{21}(Z_{i})=&\frac{Y_{i}\cdot (1-R_{i})\cdot A_{i}-E\{Y\cdot(1-R)\cdot A\}}{P(R=0,A=1)}\\
    &-\frac{E\{Y\cdot(1-R)\cdot A\}}{P(R=0,A=1)^{2}}\{(1-R_{i})\cdot A_{i}-P(R=0,A=1)\},\\
    \psi_{30}(Z_{i})=&\frac{\tilde{Y}_{i}\cdot(1- A_{i})-E\{\tilde{Y}\cdot(1- A)\}}{P(A=0)}-\frac{E\{\tilde{Y}\cdot(1-A_{i})\}}{P(A=0)^{2}}\cdot\{1-A_{i}-P(A=0)\},\\
    \psi_{31}(Z_{i})=&\frac{\tilde{Y}_{i}\cdot A_{i}-E(\tilde{Y}\cdot A)}{P(A=1)}-\frac{E(\tilde{Y}\cdot A)}{P(A=1)^{2}}\cdot\{A_{i}-P(A=1)\}.
\end{align*}
The influence functions enable estimation of the joint asymptotic variance $\Sigma$ as outlined in Section \ref{sec:setup}, which, in turn, allows us to calculate test statistics and p-values as outlined in Sections \ref{sec:testing} and \ref{sec:asympgen}. 

For sample sizes $n=200,500,1000,2000,3500$ we simulate $Z_{i}$ as follows:

\begin{align*}
    &A_{i}\sim \operatorname{Bernoulli}(1,\tfrac{1}{2}),\\
        &T_{i}\sim \operatorname{Exp}(\lambda),\\
    &R_{i}=I(T_{i}\leq\tau),\\
    &Y_{i} | R_{i}=0\sim \mathcal{N}(\mu,\sigma^{2}).
\end{align*}
We fix $\mu=40, 45$, $\sigma=15, \tau=2$ and consider the scenarios $\lambda=0.05,0.08$ corresponding to a 10\% ($\lambda=0.05$) and 15\% ($\lambda=0.08$) risk of occurrence of terminal event before landmark. The artificial score value $\Gamma$ is fixed at 15 mimicking a poor score value. 

For each configuration of sample size, $\mu$, and $\lambda$ we simulate 10,000 data sets. For each data set we calculate the generalized signed Wald test and corresponding p-value for the intersection of the three superiority hypotheses as detailed in Section \ref{sec:asympgen} based on 10,000 simulated realizations of the null-distribution. Similarly, we calculate the generalized signed Wald test for the 3 pairwise intersection tests.
We calculate type 1 errors as the fraction of p-values below a significance level $\alpha=0.025$. For the intersection of all three hypothesis we consider two sets of weights $w_{j}=\frac{1}{3}, \: j=1,2,3$ or $w_{1}=0.2, w_{2}=0.4, w_{3}=0.4$. For the pairwise intersection hypotheses  we enforce relative importance of the three hypothesis weights by using weights $\frac{1}{2},\frac{1}{2}$ for all 3  pairwise intersection tests for $w_{j}=\frac{1}{3},\: j=1,2,3$. When  $w_{1}=0.2, w_{2}=0.4, w_{3}=0.4$ we use weights $\frac{0.2}{0.2+0.4}=\frac{1}{3},\frac{0.4}{0.2+0.4}=\frac{2}{3}$ for testing $H_{1}\cap H_{2}$ and  $H_{1}\cap H_{3}$ whereas $H_{2}\cap H_{3}$ is tested using weights $\frac{0.4}{0.4+0.4}=\frac{1}{2},\frac{0.4}{0.4+0.4}=\frac{1}{2}$. Note that the weights for testing  $H_{2}\cap H_{3}$ are not varied across the two weighting scenarios. We therefore only present results concerning this intersection hypothesis for the first weighting scenario. Results are presented in Table \ref{type1sims}. 

\begin{table}
\centering
\caption{Type 1 error for testing intersection hypotheses at a nominal significance level $\alpha=0.025$. \label{type1sims} }

\begin{tabular}{rrr|rrrr}
\toprule
\multicolumn{3}{c}{}&\multicolumn{2}{c}{$w_{1}=w_{2}=w_{3}=\frac{1}{3}$} &\multicolumn{2}{c}{$w_{1}=0.2, w_{2}=w_{3}=0.4$}\\
\multicolumn{1}{r}{Hypothesis}&\multicolumn{1}{r}{Sample size}&\multicolumn{1}{r}{$\mu$}&$\lambda=0.05$ &$\lambda=0.08$&$\lambda=0.05$ &$\lambda=0.08$\\
\midrule
$H_{1}\cap H_{2}\cap H_{3}$
& 200 & 40 & 0.0260 & 0.0256 & 0.0242 & 0.0259\\
& 500 & 40 & 0.0220 & 0.0264 & 0.0239 & 0.0267\\
& 1000 & 40 & 0.0238 & 0.0259 & 0.0223 & 0.0259\\
& 2000 & 40 & 0.0265 & 0.0238 & 0.0257 & 0.0254\\
& 3500 & 40 & 0.0266 & 0.0227 & 0.0263 & 0.0242\\
& 200 & 45 & 0.0255 & 0.0252 & 0.0262 & 0.0240\\
& 500 & 45 & 0.0231 & 0.0246 & 0.0235 & 0.0246\\
& 1000 & 45 & 0.0261 & 0.0261 & 0.0255 & 0.0277\\
& 2000 & 45 & 0.0247 & 0.0240 & 0.0256 & 0.0255\\
& 3500 & 45 & 0.0238 & 0.0269 & 0.0247 & 0.0272\\

\hline
$H_{1}\cap H_{2}$

& 200 & 40 & 0.0262 & 0.0261 & 0.0249 & 0.0268\\
& 500 & 40 & 0.0218 & 0.0261 & 0.0239 & 0.0260\\
& 1000 & 40 & 0.0238 & 0.0258 & 0.0218 & 0.0250\\
& 2000 & 40 & 0.0260 & 0.0242 & 0.0261 & 0.0267\\
& 3500 & 40 & 0.0267 & 0.0225 & 0.0265 & 0.0238\\
& 200 & 45 & 0.0249 & 0.0251 & 0.0274 & 0.0220\\
& 500 & 45 & 0.0233 & 0.0259 & 0.0242 & 0.0264\\
& 1000 & 45 & 0.0261 & 0.0262 & 0.0250 & 0.0280\\
& 2000 & 45 & 0.0251 & 0.0247 & 0.0268 & 0.0264\\
& 3500 & 45 & 0.0236 & 0.0276 & 0.0236 & 0.0256\\

\hline
$H_{1}\cap H_{3}$

& 200 & 40 & 0.0238 & 0.0253 & 0.0238 & 0.0254\\
& 500 & 40 & 0.0225 & 0.0274 & 0.0241 & 0.0283\\
& 1000 & 40 & 0.0249 & 0.0270 & 0.0229 & 0.0269\\
& 2000 & 40 & 0.0256 & 0.0234 & 0.0246 & 0.0234\\
& 3500 & 40 & 0.0260 & 0.0237 & 0.0260 & 0.0254\\
& 200 & 45 & 0.0238 & 0.0276 & 0.0256 & 0.0239\\
& 500 & 45 & 0.0246 & 0.0254 & 0.0244 & 0.0255\\
& 1000 & 45 & 0.0259 & 0.0266 & 0.0240 & 0.0286\\
& 2000 & 45 & 0.0246 & 0.0256 & 0.0249 & 0.0261\\
& 3500 & 45 & 0.0238 & 0.0273 & 0.0251 & 0.0269\\

\hline
$H_{2}\cap H_{3}$
& 200 & 40 & 0.0240 & 0.0257 & - & -\\
& 500 & 40 & 0.0239 & 0.0271 & - & -\\
& 1000 & 40 & 0.0219 & 0.0260 & - & -\\
& 2000 & 40 & 0.0266 & 0.0255 & - & -\\
& 3500 & 40 & 0.0264 & 0.0239 & - & -\\
& 200 & 45 & 0.0263 & 0.0238 & - & -\\
& 500 & 45 & 0.0237 & 0.0254 & - & -\\
& 1000 & 45 & 0.0254 & 0.0272 & - & -\\
& 2000 & 45 & 0.0255 & 0.0261 & - & -\\
& 3500 & 45 & 0.0243 & 0.0278 & - & -\\

\bottomrule
\end{tabular}
\end{table}

From Table \ref{type1sims} we observe that the type 1 error is well controlled at the significance level in all scenarios.    

\begin{figure}[htpb]
  \centering
  \includegraphics[width=0.4\textwidth]{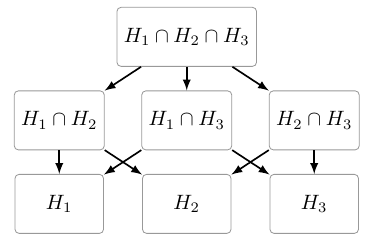}
  \caption{Illustration of a closed testing strategy for testing $H_{j},\:j=1,2,3$. A hypothesis is rejected at significance level $\alpha$ if the hypothesis itself and all hypotheses with arrows to the hypothesis are rejected at significance level $\alpha$.}
  \label{fig:closedtesting}
\end{figure}

In the third simulation study we investigate the performance of the proposed intersection tests by their combined power to reject subsets of the hypotheses $H_{j}, j=1,2,3$ when adjusting for multiple testing via a closed testing strategy \citep{closedtesting} as depicted in Figure \ref{fig:closedtesting}.

We extend the simulation setup from the second simulation study by allowing for treatment effects as follows.  For sample sizes $n=200,500,1000,2000, 3500$ we simulate $Z_{i}$ as follows:

\begin{align*}
    &A_{i}\sim \operatorname{Bernoulli}(1,\tfrac{1}{2}),\\
    &T_{i}\sim \operatorname{Exp}(\lambda+\beta_{T}\cdot A_{i}),\\
    &R_{i}=I(T_{i}\leq\tau),\\
    &Y_{i} | R_{i}=0\sim \mathcal{N}(\mu+\beta_{Y|R}\cdot A_{i},\sigma^{2}).
\end{align*}
We fix $\mu=40$, $\sigma=15, \tau=2$, $\lambda=0.07$. The parameter $\beta_{T}$ is fixed at $-0.018$ corresponding to $\theta_{1}=0.032$ and $\beta_{Y|R}$ is fixed at the value $2.7$ corresponding to $\theta_{2}=2.7$. The artificial score value $\Gamma$ is fixed at 15 mimicking a poor score value and resulting in $\theta_{3}=3.23$. These effect sizes mirror the magnitude of the estimated effects in the FLOW study.

We also note that for this specific scenario $\Sigma$ can be calculated theoretically and is given by: 

\begin{equation*}
\Sigma=\sum_{a=0}^{1}\begin{pmatrix} \frac{V(R|A=a)}{P(A=a)} & 0 & \frac{(E(Y|R=0,A=a)-\Gamma)\cdot V(R|A=a)}{P(A=a)}\\
0 & \frac{V(Y|R=0, A=a)}{P(R=0,A=a)} & \frac{V(Y| R=0, A=a)}{P(A=a)}\\
\frac{(E(Y|R=0,A=a)-\Gamma)\cdot V(R|A=a)}{P(A=a)} & \frac{V(Y| R=0, A=a)}{P(A=a)} & \frac{(E(Y|R=0,A=a)-\Gamma)^{2}\cdot V(R|A=a)}{P(A=a)}+\frac{P(R=0|A=a)\cdot V(Y|R=0, A=a)}{P(A=a)}
\end{pmatrix}
\end{equation*}

Consequently, with the configuration in this simulation study we have

\begin{equation*}
\Sigma=\begin{pmatrix} 0.405 &   0 &   10.610\\
0 & 1016.943 &  900\\
10.610 &  900 & 1075.336\\
\end{pmatrix}.
\end{equation*}
Thus we expect a close to zero correlation between $\hat{\theta}_{1}$ and $\hat{\theta}_{2}$, and, based on the first simulation study, we therefore anticipate a gain in power when applying the signed Wald test instead of the minimum p-value test in the closed testing procedure in this setting.

For each sample size, we simulate 10,000 data sets. For each data set we calculate generalized signed Wald tests based on either the equal weights or up-weighting $H_{2}$ and $H_{3}$ as in simulation study 2. We also calculate minimum p-value tests and their p-values for each intersection of hypotheses. We finally calculate the marginal p-values based on standard one-sided Wald tests to test each of the hypotheses $H_{j}, j=1,2,3$ separately. For each of the three resulting testing strategies, we evaluate $H_{j}, j=1,2,3$ according to the closed testing procedure in Figure \ref{fig:closedtesting} using a nominal significance level $\alpha=2.5\%$. Table \ref{powersims} summarizes the power of the three testing strategies to reject any specific subset of the three hypotheses as the proportion of simulated data sets in which this is achieved.

\begin{table}
  \footnotesize
\centering
\caption{Power to reject subsets of $\{H_{1},H_{2},H_{3}\}$ using each of the proposed closed testing strategies with a nominal significance level $\alpha=0.025$. \label{powersims} }

\begin{tabular}{rlrrrrrrr}
\toprule
Sample size & Test strategy & $\{H_{1}\}$ & $\{H_{2}\}$ & $\{H_{3}\}$ & $\{H_{1}, H_{2}\}$ & $\{H_{1}, H_{3}\}$ & $\{H_{2}, H_{3}\}$ & $\{H_{1}, H_{2},H_{3}\}$\\
\midrule

200 & Equal weights & 0.0841 & 0.1692 & 0.1996 & 0.0250 & 0.0618 & 0.1456 & 0.0250\\
200 & Up-weighted $H_{2},H_{3}$ & 0.0408 & 0.2033 & 0.2465 & 0.0250 & 0.0407 & 0.1794 & 0.0250\\
200 & Minimum p-value & 0.0678 & 0.1528 & 0.1996 & 0.0216 & 0.0473 & 0.1274 & 0.0216\\
500 & Equal weights & 0.1812 & 0.4071 & 0.4706 & 0.0920 & 0.1592 & 0.3788 & 0.0920\\
500 & Up-weighted $H_{2},H_{3}$ & 0.1264 & 0.4502 & 0.5371 & 0.0920 & 0.1264 & 0.4268 & 0.0920\\

500 & Minimum p-value & 0.1503 & 0.3693 & 0.4705 & 0.0835 & 0.1324 & 0.3414 & 0.0835\\
1000 & Equal weights & 0.3500 & 0.7291 & 0.8055 & 0.2727 & 0.3403 & 0.7152 & 0.2727\\
1000 & Up-weighted $H_{2},H_{3}$ & 0.3169 & 0.7529 & 0.8438 & 0.2727 & 0.3169 & 0.7423 & 0.2727\\
1000 & Minimum p-value & 0.3211 & 0.6871 & 0.8027 & 0.2624 & 0.3132 & 0.6735 & 0.2624\\
2000 & Equal weights & 0.5998 & 0.9578 & 0.9807 & 0.5774 & 0.5986 & 0.9556 & 0.5774\\

2000 & Up-weighted $H_{2},H_{3}$ & 0.5934 & 0.9617 & 0.9875 & 0.5774 & 0.5934 & 0.9608 & 0.5774\\
2000 & Minimum p-value & 0.5938 & 0.9499 & 0.9802 & 0.5750 & 0.5929 & 0.9474 & 0.5750\\
3500 & Equal weights & 0.8421 & 0.9991 & 1.0000 & 0.8414 & 0.8421 & 0.9991 & 0.8414\\
3500 & Up-weighted $H_{2},H_{3}$ & 0.8420 & 0.9991 & 1.0000 & 0.8414 & 0.8420 & 0.9991 & 0.8414\\
3500 & Minimum p-value & 0.8417 & 0.9986 & 1.0000 & 0.8411 & 0.8417 & 0.9986 & 0.8411\\
\bottomrule

\end{tabular}

\end{table}

Table \ref{powersims} shows better power performance for the equal weights generalized signed Wald test strategy when compared to the minimum p-value test strategy. This is in line with our initial expectation based on a large sample correlation of zero between $\hat{\theta}_{1}$ and $\hat{\theta}_{2}$. We also note that, compared to the equal weights strategy, the strategy that up-weighs $H_{2}$ and $H_{3}$ boosts the power to reject these two hypotheses but results in a lower power to reject $H_{1}$. We conclude that the proposed weighting in the generalized signed Wald tests can indeed function as a tool to enforce relative importance of hypotheses.

\section{Application}\label{sec:flowana}

The FLOW (Evaluate Renal Function with Semaglutide Once Weekly) clinical kidney outcome trial randomised 3,533 patients 1:1 to receive either placebo or semaglutide on top of standard of care \citep{Flow2024}. All patients had type 2 diabetes and had high-risk chronic kidney disease. High risk kidney disease patients were selected according to the estimated glomerular filtration rate (eGFR) per serum creatinine and urinary albumin to creatinine ratio (UACR). The trial duration was 5 years with a median follow-up time of 3.4 years. 

The time to first major kidney disease event \citep{Flow2023} or death from other causes define the onset of terminal event. A lower risk of having a terminal event two years after randomization corresponds to a beneficial effect of treatment. We therefore consider the null hypothesis

$$H_{1}: \theta_{1}\leq0,$$ 
where $\theta_{1}$ is the difference in probability of no terminal event between the two treatment arms at landmark year 2 after randomization. 

The eGFR measurement at landmark year 2 after randomization constitutes a surrogate marker of disease progression. A higher eGFR is indicative of a better renal function (\cite{stevens2024kdigo}). We therefore consider the null hypothesis:
$$H_{2}: \theta_{2}\leq0,$$ 
where $\theta_{2}$ is the difference in expected eGFR value between the two treatment arms given that no terminal event occurred. 

Finally, an overall risk benefit is evaluated by equating the occurrence of a
terminal event before landmark to the unfavorable eGFR value 15. This value constitutes a natural choice as current guidelines recommend renal replacement therapy for sustained eGFR values below this treshold \cite{stevens2024kdigo}. We evaluate the expected change in the resulting composite score due to treatment, which we denote $\theta_{3}$, and we note that a positive change is interpreted as a favorable risk benefit. We therefore also consider the null hypothesis

$$
H_{3}: \theta_{3}\leq 0.
$$

The treatment effects $\theta_1$ and $\theta_{2}$ are estimated as in \cite{holst2025} and $\theta_{3}$ is estimated by plugin as also outlined in \cite{holst2025}. The resulting estimates are 
\begin{align*}
    &\hat{\theta}_{1}=0.0315,\\
    &\hat{\theta}_{2}=2.681,\\
    &\hat{\theta}_{3}=3.153
\end{align*}
with an estimated asymptotic variance covariance matrix

\begin{equation*}
\hat{\Sigma}=\begin{pmatrix}
0.403 &  -1.560&  4.480\\
-1.560 & 846.241 & 822.862\\
4.480 & 822.862 & 890.064\\
\end{pmatrix}.
\end{equation*}
We note that since simulation study 3 was constructed to emulate the FLOW results the estimated effect sizes and the estimated variance covariance matrix are comparable. Based on the high rejection rates at sample size 3500 in simulation study 3,  we would up front expect that, for the FLOW trial, all three hypotheses are rejected by all three closed testing strategies employed in the simulation study. This is indeed also the case for the FLOW study where both the equal weight, up-weighted $H_{2},H_{3}$, and minimum p-value intersection tests yield p-values$<0.001$. The marginal p-values for testing $H_{1}, H_{2}$, and $H_{3}$ are $0.002, <0.001 $, and $ <0.001$, respectively.

\section{Discussion}\label{sec:disc}

In this paper we develop a general purpose Wald type test for testing intersections of one-sided null-hypotheses. In the process, we compromise a number of the aesthetic and computational virtues that characterize the popular sequentially rejective weighted Bonferroni procedures \citep{hommel2007,bretz2009,Bretz2011}. Below we discuss these properties and the impact when they are lacking.

The sequentially rejective weighted Bonferroni procedures are constructed as valid shortcuts to closed testing procedures that abide to the consonance principle \citep{Gabriel1969}. This effectively implies that if an intersection of null-hypotheses is rejected then at least one null-hypothesis in the intersection is rejected by the closed testing procedure. As a direct consequence $m$ null-hypotheses can effectively be tested in $m$ steps instead of $2^{m}-1$ steps with a shortcut bypassing all intersection tests that are rejected by default \citep{hommel2007}.

Moreover, rejection of an intersection test implies that at least one null-hypothesis in the intersection is false and should therefore intuitively lead to rejection of at least one null hypothesis. This is in fact ensured by the consonance principle and facilitates an intuitive interpretation of the intersection test. 

Our proposal is not guaranteed to adhere to the consonance principle which means that we cannot bypass any intersection tests to reduce the computational burden. However, our experience with the software implementation we have provided does not indicate that the added computational burden is of practical importance with the relatively small number of null-hypotheses that are typically considered in a clinical trial context. 

Also, there is a risk that our proposal leads to rejection of an intersection of null hypotheses without any of the separate null-hypotheses being rejected. In this case the intersection test looses its interpretation as an actual test for "any effect" and remains solely a part of a multiple testing procedure ensuring family-wise type 1 error control.  From our perspective consonance is clearly a nice to have property but definitely not a crucial prerequisite for ensuring reliable decision making under family wise type 1 error control. 

Despite the lack of consonance guarantee, the generalized signed
Wald test we propose presents a powerful and assumption-lean alternative for testing
intersections of one-sided null-hypotheses. Specifically, we have shown the signed Wald test is superior to the minimum p-value test in terms of Bahadur efficiency when testing two hypotheses, a finding supported by simulation studies showing better power performance, especially when the correlation between p-values is moderate or negative. Future work should focus on extending the theoretical characterization of the asymptotic null distribution for the general case involving the intersection of more than two hypotheses, potentially finding alternatives to the numerical optimization algorithms and Monte-Carlo simulations currently required for conducting the test.


\bibliographystyle{apalike}
\bibliography{ref}

\clearpage
\appendix

\definecolor{fgcolor}{rgb}{0.345, 0.345, 0.345}
\newcommand{\hlnum}[1]{\textcolor[rgb]{0.686,0.059,0.569}{#1}}%
\newcommand{\hlsng}[1]{\textcolor[rgb]{0.192,0.494,0.8}{#1}}%
\newcommand{\hlcom}[1]{\textcolor[rgb]{0.678,0.584,0.686}{\textit{#1}}}%
\newcommand{\hlopt}[1]{\textcolor[rgb]{0,0,0}{#1}}%
\newcommand{\hldef}[1]{\textcolor[rgb]{0.345,0.345,0.345}{#1}}%
\newcommand{\hlkwa}[1]{\textcolor[rgb]{0.161,0.373,0.58}{\textbf{#1}}}%
\newcommand{\hlkwb}[1]{\textcolor[rgb]{0.69,0.353,0.396}{#1}}%
\newcommand{\hlkwc}[1]{\textcolor[rgb]{0.333,0.667,0.333}{#1}}%
\newcommand{\hlkwd}[1]{\textcolor[rgb]{0.737,0.353,0.396}{\textbf{#1}}}%
\let\hlipl\hlkwb

\makeatletter
\newenvironment{kframe}{%
 \def\at@end@of@kframe{}%
 \ifinner\ifhmode%
  \def\at@end@of@kframe{\end{minipage}}%
  \begin{minipage}{\columnwidth}%
 \fi\fi%
 \def\FrameCommand##1{\hskip\@totalleftmargin \hskip-\fboxsep
 \colorbox{shadecolor}{##1}\hskip-\fboxsep
     \hskip-\linewidth \hskip-\@totalleftmargin \hskip\columnwidth}%
 \MakeFramed {\advance\hsize-\width
   \@totalleftmargin\z@ \linewidth\hsize
   \@setminipage}}%
 {\par\unskip\endMakeFramed%
 \at@end@of@kframe}
\makeatother
\definecolor{shadecolor}{rgb}{.97, .97, .97}
\definecolor{messagecolor}{rgb}{0, 0, 0}
\definecolor{warningcolor}{rgb}{1, 0, 1}
\definecolor{errorcolor}{rgb}{1, 0, 0}
\newenvironment{knitrout}{}{} 

\singlespacing

\section{Software implementation}\label{sec:software}

Installation of R package

\begin{Schunk}
\begin{Sinput}
> install.packages(c("targeted", "lava")) # targeted >= 0.6, lava >= 1.8.2
\end{Sinput}
\end{Schunk}

\subsection{Simulation setup}

\noindent Simulate RCT data according to setup in the third simulation study with treatment $A$, terminal event $R$, and outcome $Y$ only
observed when $R=0$.
\begin{Schunk}
\begin{Sinput}
> simdata<-function(n, # sample-size
+                   mu, sigma, 
+                   lambda, tau, gamma,
+                   trteff1, trteff2
+                   ) {
+   A <- rbinom(n, 1, 0.5) # Randomized treatment
+   Times <- rexp(n, rate = lambda + trteff1 * A) # Time to terminal event
+   R <- (Times <= tau) # Terminal event before landmark time
+   Y <- (1-R) * rnorm(n, mean = mu + trteff2 * A, sd = sigma) # Outcome among those still alive
+   Ytilde <- (1-R) * Y + gamma * R # Utility
+   return(data.frame(A, R, Y, Ytilde))
+ }
\end{Sinput}
\end{Schunk}

\noindent Function for estimating parameters of interest as defined in the simulation study
\begin{Schunk}
\begin{Sinput}
> est <- function(dat) {
+   e1 <- lm(I(1-R) ~ A, data=dat) |>
+     lava::estimate(keep="A", labels="theta1") # P(R=0|A=1)-P(R=0|A=0)
+   e2 <- lm(Y ~ A * R, data=dat) |>
+     lava::estimate(keep="A", labels="theta2")  # E(Y|R=0, A=1)-E(Y|R=0, A=0)
+   e3 <- lm(Ytilde ~ A, data=dat) |>
+     lava::estimate(keep="A", labels="theta3") # E(~Y|A=1)-E(~Y|A=0)
+   merge(e1, e2, e3)
+ }
\end{Sinput}
\end{Schunk}

\noindent Simulation of data and estimation of parameters in one scenario in the
third simulation study

\begin{Schunk}
\begin{Sinput}
> sim_args <- list(
+   n = 500,
+   lambda = 0.07,
+   tau = 2,
+   mu = 40,
+   trteff1 = -0.018,
+   trteff2 = 2.7,
+   sigma = 15,
+   gamma = 15
+ )
> 
> 
> set.seed(12345)
> dat <- do.call(simdata, sim_args)
> e <- est(dat)
> e
\end{Sinput}
\begin{Soutput}
       Estimate Std.Err     2.5%  97.5%  P-value
theta1  0.04515 0.02905 -0.01178 0.1021 0.120098
------                                          
theta2  2.99913 1.44337  0.17018 5.8281 0.037721
------                                          
theta3  3.83328 1.47377  0.94474 6.7218 0.009295
\end{Soutput}
\begin{Sinput}
> vcov(e) # Estimated asymptotic covariance matrix
\end{Sinput}
\begin{Soutput}
              theta1        theta2     theta3
theta1  8.437697e-04 -1.268575e-17 0.02176999
theta2 -1.268575e-17  2.083313e+00 1.83003132
theta3  2.176999e-02  1.830031e+00 2.17199074
\end{Soutput}
\end{Schunk}

\noindent Intersection superiority test for $H_{1}\cap H_{2}\cap H_{3}$ where
\begin{align*}
  H_{j}: \theta_j \leq 0, \qquad j=1,2,3,
\end{align*}
\begin{Schunk}
\begin{Sinput}
> targeted::test_intersection_sw(e)
\end{Sinput}
\begin{Soutput}

	Signed Wald Intersection Test

data:  
Intersection null hypothesis: theta =< [0, 0, 0]
w = [0.33, 0.33, 0.33]
Q = 0.75974, p-value = 0.0102
\end{Soutput}
\end{Schunk}

\noindent Closed-testing procedure by calculating all intersection hypotheses

\begin{Schunk}
\begin{Sinput}
> adj <- lava::closed_testing(
+     e,
+     test = targeted::test_intersection_sw,
+     noninf = rep(0, 3),
+     weights = rep(1, 3)/3
+ )
> adj$p.value
\end{Sinput}
\begin{Soutput}
    theta1     theta2     theta3 
0.06004916 0.01886060 0.01180000 
\end{Soutput}
\begin{Sinput}
> summary(adj)
\end{Sinput}
\begin{Soutput}
Call: lava::closed_testing(object = e, test = targeted::test_intersection_sw, 
    noninf = rep(0, 3), weights = rep(1, 3)/3)
\end{Soutput}

\begin{Soutput}
-- Adjusted p-values --
\end{Soutput}

\begin{Soutput}
         Estimate      adj.p
theta1 0.04515064 0.06004916
theta2 2.99913441 0.01886060
theta3 3.83327568 0.01180000
\end{Soutput}
\begin{Soutput}
-- Raw p-values for intersection hypotheses --
\end{Soutput}
\begin{Soutput}
1-way intersections:
  {theta1}                                 p = 0.0600
  {theta2}                                 p = 0.0189
  {theta3}                                 p = 0.0046

2-way intersections:
  {theta1, theta2}                         p = 0.0132
  {theta1, theta3}                         p = 0.0083
  {theta2, theta3}                         p = 0.0076

3-way intersections:
  {theta1, theta2, theta3}                 p = 0.0118
\end{Soutput}
\end{Schunk}

\end{document}